# The Positions, Colors, and Photometric Variability of Pluto's Small Satellites from HST Observations in 2005-2006

S.A. Stern
Space Science & Engineering Division
Southwest Research Institute
1050 Walnut Street
Boulder, CO 80302
astern@swri.edu

M.J. Mutchler
Space Telescope Science Institute
3700 San Martin Drive
Baltimore, MD 21218

H.A. Weaver
Space Department
11100 Johns Hopkins Road
Johns Hopkins Applied Physics Laboratory
Laurel, MD 20723

and

A.J. Steffl
Space Studies Department
Southwest Research Institute
1050 Walnut Street
Boulder, CO 80302

# Abstract

Pluto's two small satellites, temporarily designated S/2005 P 1 and S/2005 P 2, were observed on four dates (15.1 and 18.1 May 2005, 15.7 February 2006, and 2.8 March 2006) using the Hubble Space Telescope's (HST) Advanced Camera for Surveys (ACS). Here we collect together the astrometric positions of these two satellites (henceforth "P1" and "P2"), as well as a single color measurement for each satellite and initial constraints on their photometric variability obtained during these observations. We find that both satellites have essentially neutral (grey) reflectivities, like Charon. We also find that neither satellite exhibited strong photometric variation, which might suggest that P1 and P2 are toward the large end of their allowable size range, and therefore may have far lower reflectivities than Charon.

# 1 Introduction

The discovery of two small satellites of Pluto and the initial orbit characterization for those satellites were described by Weaver et al. (2006). Confirmation of this discovery was made by HST on 15 February 2006 (Mutchler et al. 2006). Figure 1 presents the highest resolution image of the satellite system made to date.

The implications of these discoveries for the origin of Pluto's satellite system and the possibility of associated dust rings around Pluto were discussed by Stern et al. (2006). Steffl et al. (2006) used the combined HST dataset to set stringent upper limits on the brightness of any additional satellites throughout the Pluto system, finding that any undiscovered satellites would have to be far fainter than either P1 or P2, except in the region between Pluto and Charon.

Here we present a full set of astrometric positions of the two new satellites from all existing 2005-2006 HST observations, as well as color measurements and some initial constraints on photometric variability of the two satellites that these data provide.

We point out that Buie et al. (2006) published an improved orbit for Pluto's satellites by combining the astrometry we reported in the May 2005 discovery observations with archival positions they derived from HST ACS Pluto-system images obtained in 2002 and 2003. Additionally, dynamical implications of the new satellites were discussed in recent papers by Ward & Canup (2006) and Lee & Peale (2006). Interest in the Pluto system's newly-found richness has been highlighted by Binzel (2006).

# 2 Observations and Astrometric Positions

All of our Pluto system observations of P1 (Pluto's outer small satellite) and P2 (Pluto's inner small satellite) were obtained with the HST using the Advanced Camera for Surveys (ACS). The ACS camera modes, filters, and the observing circumstances on each of our four HST observing dates are described in Table 1. The telescope was programmed to track the apparent motion of Pluto for all of these observations.

**Table 1**
**Observation Summary**

| Observing Date | ACS Mode | ACS Filter(s) | Heliocentric Distance (AU) | Geocentric Distance (AU) | Solar Phase (deg) |
|---|---|---|---|---|---|
| 15.05 May 2005 | WFC | F606W | 30.95 | 30.08 | 0.96 |
| 18.14 May 2005 | WFC | F606W | 30.95 | 30.05 | 0.88 |
| 15.66 Feb 2006 | HRC | F606W | 31.07 | 31.54 | 1.58 |
| 02.75 Mar 2006 | HRC | F606W, F435W | 31.08 | 31.31 | 1.77 |

Notes: WFC=Wide Field Channel; HRC=High Resolution Channel.

The F606W ("Broad V") ACS filter, which has a center wavelength of 591.8 nm and a width of 67.2 nm, was used on all four observing dates. To obtain color information, we additionally used the ACS F475W filter ("Johnson B"), which has a center wavelength of 429.7 nm and a width of 103.8 nm, on 2 March 2006. On all dates, unsaturated images of Pluto and Charon were obtained by using short exposure times (either 0.5, 1, or 3 s); much longer exposure times (145 or 475 s) were used to obtain deep images of the new satellites, which are roughly 5000 times fainter than Pluto itself.

Astrometric positions of P1 and P2 derived from the ACS images on each date are given in Table 2; all positions are referenced to Pluto's center of light, which can vary from Pluto's physical center, owing to albedo variegations by up to ~0.025 arcsec.

**Table 2**
**Astrometric Positions**

| Date | P1 Position Angle (deg) | P1 Distance (") | P2 Position Angle (deg) | P2 Distance (") |
|---|---|---|---|---|
| 15.05 May 2005 | 264.2±0.5 | 1.85±0.035 | 326.9±0.5 | 2.09±0.035 |
| 18.14 May 2005 | 305.8±0.5 | 2.36±0.035 | 355.5±0.5 | 2.22±0.035 |
| 15.66 Feb 2006 | 343.0±0.3 | 2.86±0.010 | 355.3±0.3 | 2.03±0.010 |
| 02.75 Mar 2006 | 138.2±0.5 | 2.69±0.015 | 213.7±0.5 | 1.43±0.015 |

Note: Errors displayed here are relative to Pluto's center of light; center of light to center-of-body corrections introduce possible additional errors of ~0.025 arcsec.

These positions, when combined with archival detections of P1 and P2 (e.g., Buie et al. 2006), and future positions, can be combined to provide improved orbital fits for Pluto's small satellites. Although beyond the scope of this paper, we point out the utility of such orbit fits (i) for encounter planning for the New Horizons Pluto system flyby which will occur in mid-2015 (Stern & Cheng 2002), (ii) for further refining the Pluto/Charon mass ratio (Buie et al. 2006), (iii) for dynamical studies which may yield masses for one or both of the small satellites (Lee & Peale 2006), and (iv) for determining whether or not P1 and P2 are in fact in mean motion resonances with either one another or Charon, or both (Weaver et al. 2006).

## 3 Photometric Variability Measurements

Accurate V-band photometry of P1 and P2 was obtained on three of the four HST observation dates (see Table 3). (It was not possible to obtain accurate photometry for P1 and P2 on one date each in May 2005 because the satellites unfortunately fell near diffraction spikes of Pluto, which prevented accurate measurements.) The V-band magnitudes shown in Table 3 were derived from the instrumental magnitudes following the prescription described by Sirianni et al. (2005).

**Table 3**
**Photometric V Magnitudes**

| Date | P1 V Magnitude | P2 V Magnitude |
|---|---|---|
| 15.05 May 2005 | N/A | 23.38±0.17 |
| 18.14 May 2005 | 22.93±0.12 | N/A |
| 15.66 Feb 2006 | 23.26±0.15 | 23.70±0.20 |
| 02.75 Mar 2006 | 23.30±TBD | 23.57±TBD |

Using the data in Tables 1 and 3, we normalized the various V magnitude measurements of P1 and P2, adjusting for the changing heliocentric and geocentric distances, and for the changing solar phase angle. For the phase angle correction, we assumed that P1 and P2 follow Charon's phase law (Tholen & Buie 1997). After making these corrections, we find that the relative brightnesses of P1 on its three observing dates are 1.00:0.78:0.73, and the values for P2 are 1.00:0.79:0.86.

The observed relative brightness changes are not large; they indicate that the relative changes in effective radius presented to us on the various observing dates fluctuated by <14%. Of course, our temporal sampling is sparse and presumably does not provide a good characterization of the lightcurve of either P1 or P2. Nonetheless, if P1 and P2 have Charon-like albedos of 0.35, their nominal radii are just 61 and 46 km, respectively (Weaver et al. 2006). Such small bodies often display radii variations of order 2:1 as they rotate, which would produce relative photometric variations of order a factor of four, depending on the pole orientation.

Why might P1 and P2 display smaller fluctuations? One explanation might be that we serendipitously observed both satellites at times when the same faces were presented to us; this is unlikely but certainly possible. Another possibility is that P1 and P2 might have albedo variations that serendipitously compensate for cross-sectional changes, another unlikely possibility.

A third possibility is that P1 and P2, in fact, have substantially lower albedos than Charon. For example, if both have V albedos of 0.04, then their radii are near 167 and 137 km, respectively (Weaver et al. 2006). Bodies of such radii are typically more round than bodies having radii <70 km. Thus, the lack of brightness variations in our data could be hinting that P1 and P2 may be rounder owing to their having lower albedos than Charon, implying they are larger than expected by assuming Charon-like albedos.

## 4 Color Measurements

On 2006 March 2.75, we obtained images of the Pluto system using both the ACS F435W and F606W filters. This allowed us to obtain simultaneous B and V photometry for P1 and P2, which in turn allows us the opportunity to obtain surface colors. Using the background-subtracted B and V count rates of each object, and adopting Charon's B-V=+0.710 (Buie et al. 1997) as a calibration reference, we derive B-V colors of +0.653±0.026 and +0.654±0.065 for P1 and P2, respectively. From these colors, we conclude that all three of Pluto's satellites display similar, essentially solar colors; this in turn implies that they all have grey intrinsic surface reflectances in the visible.

Buie et al. (2006) also found a neutral B-V color for P1 in lower-SNR measurements made by co-adding many archival Pluto-system images in 2002 and 2003. However, they found a much redder B-V=+0.91±0.15 for P2 from similar SNR data. Our results, with significantly smaller error bars, suggest that the red color suggested for P2 by Buie et al. is an artifact, and that both P1 and P2 have a B-V color similar to Charon.

This color similarity among Pluto's three satellites is not surprising given that all three bodies are relatively small, none have atmospheres, that they seem to share a common origin (Stern et al. 2006), and that all have experienced the same environmental (e.g., radiation dose) effects for many Gyr.

# Acknowledgements

Support for this work was provided by NASA through Grants GO-10427 and GO-10774 from the Space Telescope Science Institute, which is operated by the Association of Universities for Research in Astronomy, Inc., and the New Horizons Pluto-Kuiper Belt mission.

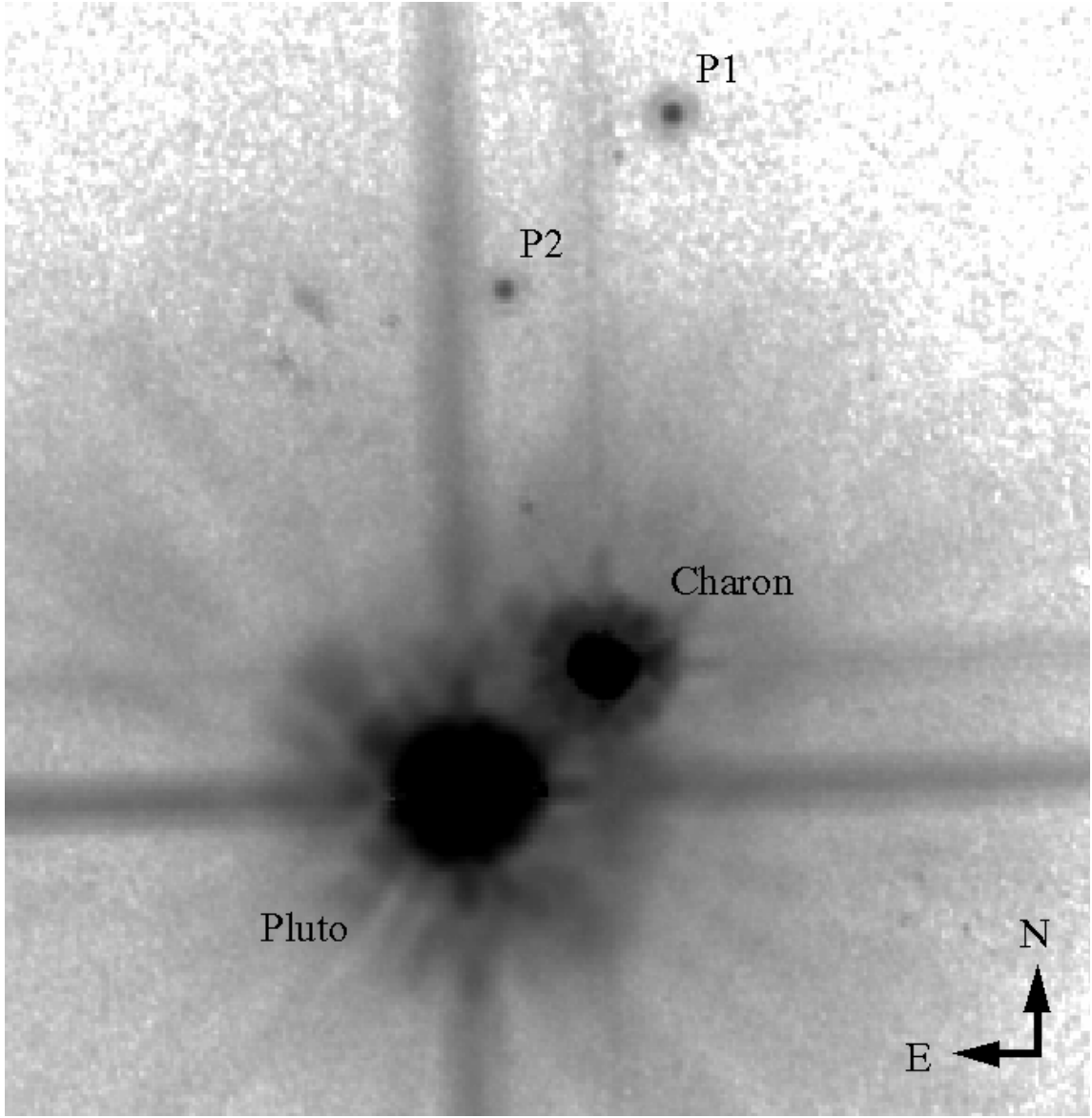

Figure 1. ACS/HRC broad V-band composite image of the Pluto system obtained on 15 February 2006. The image above is the combination of four 475s dithered images drizzled to an output scale of 0.015 arcsec/pixel (smaller than the input HRC detector pixels, which are 0.025 arcsec/pixel). This is the highest-resolution image of the entire Pluto satellite system to date. The complex HRC point spread function and diffraction spikes emanating from Pluto and Charon are evident. All other faint features can be associated with residual star trails or many cosmic rays events that were not completely removed by our processing steps.